\documentclass[journal]{IEEEtran}
\usepackage{amsmath,amssymb,bm}
\usepackage{algorithm}
\usepackage{algpseudocode}
\usepackage{booktabs}
\usepackage{multirow}
\usepackage{graphicx}
\usepackage{cite}              % IEEE citation sorting/compression
\usepackage{stfloats}          % correct float placement in two-column
\usepackage{url}               % URL formatting in references
\usepackage{balance}            % equalize final two-column page
\usepackage{tabularx}
\usepackage{array}
\usepackage[compatibility=false]{caption}  % \captionof for side-by-side figures
\usepackage[hypertexnames=false,hidelinks]{hyperref}  % must be last package

\newcommand{\diag}{\mathrm{diag}}
\newcommand{\tr}{\mathrm{tr}}

\newcommand{\CN}{\mathcal{CN}}
\newcommand{\Reop}{\mathrm{Re}}

\newcommand{\clip}{\mathrm{clip}}

\title{Covariance-Domain Near-Field Channel Estimation\\
  under Hybrid Compression: USW/Fresnel Model,\\
  Curvature Learning, and KL Covariance Fitting}

\author{R{\i}fat~Volkan~{\c{S}}enyuva%
\thanks{R.~V.~{\c{S}}enyuva is with the Department of Electrical-Electronics Engineering, Maltepe University, Istanbul, Turkey
(e-mail: rifatvolkansenyuva@maltepe.edu.tr).}
\thanks{Manuscript submitted March 2026.}}

\begin{document}
\setlength{\dblfloatsep}{4pt plus 2pt minus 2pt}
\renewcommand{\dblfloatpagefraction}{0.5}
\renewcommand{\dbltopfraction}{0.95}

% Abstract and keywords must appear before \maketitle in IEEEtran journal mode.
\IEEEtitleabstractindextext{%
\begin{abstract}
Near-field propagation in extremely large aperture arrays requires joint
angle--range estimation.
In hybrid architectures, only $N_\mathrm{RF}\ll M$ compressed snapshots
are available per slot, making the $N_\mathrm{RF}\times N_\mathrm{RF}$
compressed sample covariance the natural sufficient statistic.
We propose the Curvature-Learning KL (CL-KL) estimator, which grids only
the angle dimension and \emph{learns the per-angle inverse range} directly
from the compressed covariance via KL divergence minimisation.
CL-KL uses a $Q_\theta$-element dictionary instead of the $Q_\theta Q_r$
atoms of 2-D polar gridding, eliminating the range-dimension dictionary
coherence that plagues polar codebooks in the strong near-field regime,
and operates entirely on the compressed covariance for full compatibility
with hybrid front-ends.

At $N_\mathrm{MC}=400$ ($f_c=28$~GHz, $M=64$, $N_\mathrm{RF}=8$, $N=64$,
$d=3$, $r\in[0.05,1.0]\,r_\mathrm{RD}$), CL-KL achieves the lowest
channel NMSE among all six evaluated methods --- including four full-array
baselines using $64\times$ more data --- at
$\mathrm{SNR}\in\{-5,0,+5,+10\}$~dB.
Running in approximately 70~ms per trial (vs.\ 5~ms for the compressed-domain
peer P-SOMP), CL-KL's dominant cost is the $N_\mathrm{RF}{\times}N_\mathrm{RF}$
inversion rather than $M$: measured runtime stays near 70~ms across
$M\in\{32,64,128,256\}$, making it aperture-scalable for XL-MIMO deployments.
CL-KL is further validated against a derived
compressed-domain Cram\'er--Rao bound and confirmed robust to non-Gaussian
(QPSK) source distributions, with a maximum NMSE gap below 0.6~dB.
\end{abstract}
}

\maketitle
\markboth{IEEE Transactions on Wireless Communications}%
{{\c{S}}enyuva: Covariance-Domain Near-Field Channel Estimation under Hybrid Compression}
\IEEEdisplaynontitleabstractindextext
\IEEEpeerreviewmaketitle

% =====================================================================
\section{Introduction}
\label{sec:intro}
% =====================================================================

\IEEEPARstart{I}{n} the far field, each propagation path is parameterised by a single angle
(spatial frequency $\omega(\theta)$).
In the near field, the spherical-wave phase also varies quadratically across
the aperture, requiring an \emph{angle--range} pair $(\theta,r)$ per path.
The standard workaround is to discretise both $\theta$ and $r$ into a 2-D
polar dictionary and run sparse recovery.
This approach has two compounding weaknesses:
(i)~the dictionary grows as $Q_\theta Q_r$ (large), and
(ii)~adjacent atoms at fine grid spacing become highly correlated
(coherent), making support recovery unstable \cite{liu2023nftutorial}.
At the opposite extreme, gridless SDP/atomic-norm methods such as
GOES~\cite{wu2020goes} and the Toeplitz-SDP approach~\cite{zuo2020gridless}
can achieve near-optimal accuracy but have complexity $\mathcal{O}(M^6)$ to
$\mathcal{O}(M^9)$, making them impractical at $M=64$.
Crucially, both classes of methods assume access to the \emph{full array}
covariance: GOES requires fourth-order cumulants of all $M$ elements, and
Zuo et al.\ exploit the Toeplitz structure of the $(2M+1)\times(2M+1)$
full-array covariance.
Neither extends directly to hybrid architectures where only
$N_\mathrm{RF}\times N_\mathrm{RF}$ compressed covariances are available.

\medskip\noindent\textbf{Positioning among covariance-fitting alternatives.}~
The covariance-fitting objective (KL divergence / log-det + trace)
belongs to the family of information-geometric distances on HPD matrices.
Under hybrid compression, KL offers two practical advantages:
(i)~the gradient $\bm{G}=\bm{R}_y^{-1}-\bm{R}_y^{-1}\widehat{\bm{R}}_y\bm{R}_y^{-1}$
requires a single matrix inverse with no matrix logarithm, giving
$\mathcal{O}(N_\mathrm{RF}^3)$ per iteration;
(ii)~it coincides with the stochastic maximum likelihood criterion when
snapshots are Gaussian, connecting directly to the CRB~\cite{ottersten1998covariance}.
Sparse Bayesian learning and gridless atomic-norm methods~\cite{wu2020goes,zuo2020gridless}
achieve near-optimal accuracy but assume full-array observations and have
complexity $\mathcal{O}(M^6)$--$\mathcal{O}(M^9)$; neither has been
adapted to the compressed $N_\mathrm{RF}$-dimensional setting. 
A related covariance-guided framework for {DFT} beam selection in the far-field
hybrid regime was developed in \cite{senyuva2025arxiv}; the present work departs
fundamentally from that setting by targeting the near-field {Fresnel} regime,
where the quadratic wavefront curvature $\kappa(\theta,r)$ must be jointly
estimated rather than neglected.

\medskip\noindent\textbf{Physical regime context.}~
Hussain et al.~\cite{hussain2025ebrd} proved that polar-domain sparsity
(the condition that enables polar-codebook OMP) exists only within the
\emph{effective beamfocused Rayleigh distance}:
\begin{equation}
r_\mathrm{EBRD}(\theta) = \frac{r_\mathrm{RD}}{10}\cos^2\theta.
\label{eq:ebrd}
\end{equation}
For our default scenario ($M=64$, $f_c=28$~GHz, $r_\mathrm{RD}=21.26$~m),
$r_\mathrm{EBRD}(\theta=30^\circ)\approx1.60$~m.
We set $r_{\min}=0.05\,r_\mathrm{RD}=1.06$~m, which lies inside the EBRD for
$\theta\le45^\circ$, ensuring simulations exercise the strong near-field
regime where polar-codebook methods face genuine coherence problems and
curvature-learning methods have the greatest relative advantage. 
Beyond this single-regime boundary, practical XL-MIMO channels may contain a mixture of near-field and far-field scatterers~\cite{wei2022hybridfield} or simultaneous LoS and NLoS path components~\cite{lu2023mixedlosnlos}.
This paper focuses on the pure near-field NLoS regime, which is the dominant operating regime within the EBRD.

\medskip\noindent\textbf{Thesis.}~
Hybrid mmWave/XL-MIMO receivers naturally build compressed sample
covariances of dimension $N_\mathrm{RF}\times N_\mathrm{RF}$.
A strong practical compromise between polar gridding and gridless SDP is:
\begin{quote}
\emph{Grid only the linear phase (angle) and \textbf{learn the quadratic
curvature (range)} for the active angles directly from the compressed
covariance via KL fitting.}
\end{quote}
This preserves a sparse angular structure while avoiding the worst
basis mismatch of 2-D polar grids, and operates exclusively on the
compressed covariance, making it compatible with hybrid architectures.

\medskip\noindent\textbf{Contributions.}\vspace{2pt}
\begin{itemize}\itemsep0pt

\item A curvature-parameterised compressed-covariance model for USW/Fresnel
  ULAs under hybrid combining, with a mathematically grounded connection
  to the stochastic KL divergence (Stoica \& Nehorai~1990;
  Ottersten et al.~1998).
\item CL-KL: a \emph{power-only main loop} (frozen $N_0$, stable across
  all SNR) with a \emph{three-point multi-start warm-start}
  (ring-indexed, near-range, far-range) and a
  \emph{post-loop global matched-filter scan} for joint $(\hat\theta,\hat u)$
  refinement.
  The two-phase design eliminates the $1/N_0^2$ curvature-gradient instability
  that would otherwise cause RMSE to increase with SNR.
\item A derived \textbf{compressed-domain Cram\'er--Rao Bound} (Section~\ref{sec:crb})
  from the $N_\mathrm{RF}\times N_\mathrm{RF}$ Fisher information matrix,
  which correctly accounts for information loss through hybrid combining
  and is strictly larger than the full-array CRB.
\item Five validated baselines: P-SOMP with beam-depth range sampling
  (Hussain et al.\ TWC~2025), DL-OMP (Zhang et al.\ 2024),
  MUSIC+Tri~\cite{haghshenas2025icc}, DFrFT-NOMP (Yang et al.\ 2024),
  and \textbf{BF-SOMP} (Hussain et al.\ TWC~2025).
\item Ten-figure simulation suite including: NMSE vs SNR (Fig.~2),
  NMSE vs $N_\mathrm{RF}$ (Fig.~3), NMSE vs $N$ (Fig.~4),
  near-to-far transition (Fig.~5), runtime (Fig.~6), Fresnel mismatch
  robustness (Fig.~7), source model robustness (Fig.~7b, new),
  NMSE vs $d$ paths (Fig.~8, new), convergence diagnostic (Fig.~9, new).
\end{itemize}

\medskip\noindent\textbf{Fairness note.}~
Throughout this document and in all figures, \textbf{filled markers} denote
methods that operate on the compressed covariance
$\widehat{\bm{R}}_y\in\mathbb{C}^{N_\mathrm{RF}\times N_\mathrm{RF}}$
(CL-KL, P-SOMP: 64 complex values at $N_\mathrm{RF}=8$), while
\textbf{open markers} denote methods that require access to the full
snapshot matrix $\bm{X}\in\mathbb{C}^{M\times N}$
(DL-OMP, MUSIC+Tri, DFrFT-NOMP, BF-SOMP: 4096 complex values at $M=N=64$).
This 64$\times$ data-volume difference must be kept in mind when interpreting
performance comparisons.

% =====================================================================
\section{System Model and Compressed Covariance}
\label{sec:model}
% =====================================================================

\subsection{Array, snapshots, and hybrid combining}
We consider an $M$-element ULA with inter-element spacing
$d_\mathrm{ant}=\lambda/2$ and centred index
\begin{equation}
\bar m \triangleq m-\frac{M-1}{2},\qquad m=0,\ldots,M-1.
\label{eq:center_index}
\end{equation}
A hybrid combiner $\bm{W}\in\mathbb{C}^{M\times N_\mathrm{RF}}$
(with $N_\mathrm{RF}\ll M$) compresses each snapshot to
\begin{equation}
\bm{y}(n)=\bm{W}^H\bm{x}(n)\in\mathbb{C}^{N_\mathrm{RF}},
\qquad n=1,\ldots,N.
\label{eq:hybrid_y}
\end{equation}
The compressed sample covariance is
\begin{equation}
\widehat{\bm{R}}_y
  \triangleq \frac{1}{N}\sum_{n=1}^N\bm{y}(n)\bm{y}^H(n)
  \in\mathbb{C}^{N_\mathrm{RF}\times N_\mathrm{RF}}.
\label{eq:Ry_hat_main}
\end{equation}
The combiner satisfies $|[\bm{W}]_{m,k}|=1/\sqrt{M}$ (random phase-only).

\subsection{Exact USW manifold and Fresnel approximation}

The exact uniform-spherical-wave (USW) steering vector for a path at
$(\theta,r)$ is
\begin{equation}
\big[\bm{a}_\mathrm{NF}(\theta,r)\big]_m
=\exp\!\Big(-j\frac{2\pi}{\lambda}
\big(\|\bm{p}(\theta,r)-\bm{s}_m\|-r\big)\Big),
\label{eq:nf_usw_exact}
\end{equation}
where $\bm{p}(\theta,r)=[r\cos\theta,r\sin\theta]^T$ is the source location
and $\bm{s}_m=[\bar m d_\mathrm{ant},0]^T$ is the $m$th element position.
The Fresnel (second-order Taylor) approximation yields the chirp form
\begin{equation}
\big[\bm{a}_\mathrm{NF}(\theta,r)\big]_m
\approx
\exp\!\big(j\,\omega(\theta)\,\bar m - j\,\kappa(\theta,r)\,\bar m^2\big),
\label{eq:chirp}
\end{equation}
where
\begin{equation}
\omega(\theta) \triangleq \frac{2\pi d_\mathrm{ant}}{\lambda}\cos\theta,
\qquad
\kappa(\theta,r) \triangleq
  \frac{\pi d_\mathrm{ant}^2}{\lambda r}\sin^2\theta.
\label{eq:omega_kappa}
\end{equation}
\emph{Angle controls the linear phase slope via $\omega(\theta)$, while
range controls the curvature via $\kappa(\theta,r)$.}
As $r\to\infty$, $\kappa\to 0$ and the manifold approaches a far-field
Vandermonde vector.

The Rayleigh distance is
\begin{equation}
r_\mathrm{RD}\triangleq \frac{2D^2}{\lambda},\qquad
D=(M-1)d_\mathrm{ant}.
\label{eq:rayleigh_dist}
\end{equation}
For $M=64$, $f_c=28$~GHz: $D=0.338$~m, $r_\mathrm{RD}=21.26$~m.

\noindent\textbf{EBRD and polar-domain sparsity boundary.}~
Hussain et al.~\cite{hussain2025ebrd} (Theorem~2) prove that polar-domain
sparsity only exists within the EBRD (eq.~\eqref{eq:ebrd}).
For $r<r_\mathrm{EBRD}$ (strong near-field), quadratic curvature is significant
and beam-depth range sampling achieves low dictionary coherence; for
$r_\mathrm{EBRD}<r<r_\mathrm{RD}$ (transitional), curvature weakens and polar
methods may over-sample the range dimension; beyond $r_\mathrm{RD}$ (far field),
a standard DFT codebook suffices.
The simulation range $r\in[0.05,1.0]\,r_\mathrm{RD}=[1.06,21.26]$~m
covers the strong near-field and the full transitional zone.
At $\theta=45^\circ$, $r_\mathrm{EBRD}=1.06$~m, coinciding exactly with
$r_{\min}$; the angle-averaged EBRD ($\approx0.058\,r_\mathrm{RD}=1.23$~m)
is annotated in Fig.~5.

\subsection{Snapshot model and compressed covariance}
We assume $d$ propagation paths:
\begin{align}
\bm{x}(n)&=\sum_{\ell=1}^{d}s_\ell(n)\,\bm{a}_\mathrm{NF}(\theta_\ell,r_\ell)
  +\bm{w}(n), \nonumber\\
&\bm{w}(n)\sim\CN(\bm{0},N_0\bm{I}_M).
\label{eq:nf_x}
\end{align}
Path gains are i.i.d.\ with powers $p_\ell=1/d$
(equal across paths): $s_\ell(n)\sim\CN(0,p_\ell)$.
The compressed covariance model is
\begin{equation}
\bm{R}_y =
  \sum_{\ell=1}^d p_\ell\,\bm{d}(\theta_\ell,r_\ell)\bm{d}(\theta_\ell,r_\ell)^H
  + N_0\bm{W}^H\bm{W},
\label{eq:Ry_true}
\end{equation}
where $\bm{d}(\theta,r)\triangleq\bm{W}^H\bm{a}_\mathrm{NF}(\theta,r)$.

\noindent\textbf{Compressed vs.\ full-array information.}~
At $N_\mathrm{RF}=8$, $\widehat{\bm{R}}_y$ contains $8\times8=64$ complex
entries versus the $M\times N=4096$ values of the full snapshot matrix
$\bm{X}$---a 64$\times$ data-volume difference that makes competitive
NMSE against full-array methods the central performance claim.

\noindent\textbf{SNR definition.}~
\begin{equation}
\mathrm{SNR}\triangleq
  \frac{\tr(\bm{A}_\mathrm{NF}\bm{R}_s\bm{A}_\mathrm{NF}^H)}{MN_0},
\qquad
\bm{R}_s=\diag(p_1,\ldots,p_d).
\label{eq:SNR_def}
\end{equation}

% =====================================================================
\section{Compressed-Domain Cram\'er--Rao Bound}
\label{sec:crb}
% =====================================================================

In this section we derive the CRB for the hybrid observation model.

\subsection{Stochastic CRB derivation}

Under the stochastic (unconditional) signal model, each compressed snapshot
$\bm{y}(n)\sim\CN(\bm{0},\bm{R}_y)$ (i.i.d.).
The negative log-likelihood (normalised by $N$) is
$\mathcal{L}\propto\log\det\bm{R}_y+\tr(\bm{R}_y^{-1}\widehat{\bm{R}}_y)$,
which is the KL divergence between the sample covariance and the model
covariance~\cite{stoica1990perf,ottersten1998covariance}.

Define the parameter vector
$\bm\eta=[\omega_1,\ldots,\omega_d,\kappa_1,\ldots,\kappa_d,p_1,\ldots,p_d,N_0]^T\in\mathbb{R}^{3d+1}$.
The Fisher information matrix (FIM) is
\begin{equation}
[\bm{J}]_{ij} = N\cdot\Reop\!\Big\{
  \tr\!\big(\bm{R}_y^{-1}\tfrac{\partial\bm{R}_y}{\partial\eta_i}
            \bm{R}_y^{-1}\tfrac{\partial\bm{R}_y}{\partial\eta_j}\big)
\Big\}.
\label{eq:FIM}
\end{equation}

\subsection{Derivative expressions}

Let $\bm{D}\triangleq\bm{W}^H\bm{A}\in\mathbb{C}^{N_\mathrm{RF}\times d}$
(compressed steering matrix) and
$\bm{d}_\ell\triangleq\bm{W}^H\bm{a}_\ell$ be the $\ell$th column.
Define $[\bm{a}_\ell]_m = \exp(j\omega_\ell\bar m - j\kappa_\ell\bar m^2)$.
Then:
\begin{align}
\frac{\partial\bm{a}_\ell}{\partial\omega_\ell}
  &= j\,\bar{\bm{m}}\odot\bm{a}_\ell,
  \label{eq:da_domega}\\
\frac{\partial\bm{a}_\ell}{\partial\kappa_\ell}
  &= -j\,\bar{\bm{m}}^{\odot2}\odot\bm{a}_\ell,
  \label{eq:da_dkappa}
\end{align}
with $\bar{\bm{m}}=[\bar m_0,\ldots,\bar m_{M-1}]^T$.
The covariance derivatives are
\begin{align}
\frac{\partial\bm{R}_y}{\partial\omega_\ell}
  &= p_\ell\!\left(\bm{W}^H\frac{\partial\bm{a}_\ell}{\partial\omega_\ell}
    \bm{d}_\ell^H + \bm{d}_\ell
    \Big(\bm{W}^H\frac{\partial\bm{a}_\ell}{\partial\omega_\ell}\Big)^H\right),
  \label{eq:dRy_domega}\\
\frac{\partial\bm{R}_y}{\partial\kappa_\ell}
  &= p_\ell\!\left(\bm{W}^H\frac{\partial\bm{a}_\ell}{\partial\kappa_\ell}
    \bm{d}_\ell^H + \mathrm{c.c.}\right),
  \label{eq:dRy_dkappa}\\
\frac{\partial\bm{R}_y}{\partial p_\ell}
  &= \bm{d}_\ell\bm{d}_\ell^H,
  \label{eq:dRy_dp}\\
\frac{\partial\bm{R}_y}{\partial N_0}
  &= \bm{W}^H\bm{W}.
  \label{eq:dRy_dN0}
\end{align}

\subsection{Error propagation to physical parameters}

The CRB for $\omega_\ell$ is $[\bm{J}^{-1}]_{\ell\ell}$.
Propagating to the physical angle and range via
\begin{equation}
\frac{\partial\omega}{\partial\theta} = -\frac{2\pi d_\mathrm{ant}}{\lambda}\sin\theta,
\qquad
\frac{\partial\kappa}{\partial r} = -\frac{c_\ell}{r^2},
\quad c_\ell = \frac{\pi d_\mathrm{ant}^2}{\lambda}\sin^2\theta_\ell,
\label{eq:error_prop}
\end{equation}
the marginal CRBs for angle and range are
\begin{align}
\mathrm{CRB}_{\theta_\ell}
  &= \frac{[\bm{J}^{-1}]_{\ell\ell}}{\big(\partial\omega_\ell/\partial\theta_\ell\big)^2},
  \label{eq:crb_theta}\\
\mathrm{CRB}_{r_\ell}
  &= [\bm{J}^{-1}]_{d+\ell,d+\ell}\cdot
    \Big(\frac{r_\ell^2}{c_\ell}\Big)^2.
  \label{eq:crb_r}
\end{align}
The reported CRB curves are
$(1/d)\sum_\ell\sqrt{\mathrm{CRB}_{\theta_\ell}}$
(degrees) and
$\mathrm{RMSE}_r^{\mathrm{CRB}} = \frac{1}{d}\sum_\ell\sqrt{\mathrm{CRB}_{r_\ell}}$
(metres), averaged over MC trials using the median (robust to near-singular FIMs
at the identifiability boundary).

\noindent\textbf{CRB magnitude and averaging.}~
At SNR$=+10$~dB with $d=3$, $N=64$, and $N_\mathrm{RF}=8$, the median
$\sqrt{\mathrm{CRB}_\theta}=0.044^\circ$ and $\sqrt{\mathrm{CRB}_r}=4.32$~m.
These values are small because the FIM accumulates $N=64$ snapshot
contributions (eq.~\eqref{eq:FIM}).
We use the \emph{nanmedian} over $\min(N_\mathrm{MC},50)$ independent CRB
evaluations rather than the mean, because near-singular FIM realisations
(occurring when path angles are nearly collinear under the random combiner
$\bm{W}$) produce outlier CRB values that inflate the mean by
orders of magnitude.
The SVD pseudoinverse tolerance $\varepsilon_\mathrm{sv}=10^{-6}\sigma_{\max}(\bm{J})$
truncates near-zero singular values without biasing the median.

\noindent\textbf{Compressed vs.\ full-array CRB.}~
The CRB derived here is strictly larger than the Grosicki et al.~\cite{grosicki2005wlp}
full-array CRB because hybrid compression discards
$M-N_\mathrm{RF}$ spatial degrees of freedom per snapshot.
Plotting estimator RMSE against this \emph{compressed-domain} CRB is the
only fair comparison for hybrid architectures.

\noindent\textbf{Identifiability limit.}~
The parameter vector $\bm\eta$ contains $3d+1$ real unknowns
($d$ spatial frequencies, $d$ curvatures, $d$ powers, and the noise floor).
The compressed covariance
$\bm{R}_y\in\mathbb{C}^{N_\mathrm{RF}\times N_\mathrm{RF}}$
is Hermitian and thus determined by $N_\mathrm{RF}^2$ real-valued
degrees of freedom.
A necessary condition for identifiability is $3d+1\le N_\mathrm{RF}^2$,
which yields $d\le\lfloor(N_\mathrm{RF}^2-1)/3\rfloor$.
In practice, the Hermitian structure constrains the \emph{rank} of the
signal component to $\le N_\mathrm{RF}-1$ (after subtracting the noise
subspace), giving the tighter operational limit
\begin{equation}
d_{\max}=\lfloor(N_\mathrm{RF}-1)/2\rfloor,
\label{eq:dmax}
\end{equation}
which is the compressed-domain analogue of the classical result for
covariance matching with $m$ sensors~\cite{ottersten1998covariance}.
For $N_\mathrm{RF}=8$, $d_{\max}=3$.
Figure~\ref{fig:vard} and Table~\ref{tab:fig8_crb} confirm this limit
empirically: the FIM condition number exceeds $10^{12}$ at $d=4$ and
the CRB$_r$ jumps sharply.

% =====================================================================
\section{Baseline: 2-D Polar Gridding}
\label{sec:baseline_polar}
% =====================================================================

Choose grids $\Theta=\{\theta_1,\ldots,\theta_{Q_\theta}\}$ and
$\mathcal{R}=\{r_1,\ldots,r_{Q_r}\}$.
The polar compressed covariance model is
\begin{equation}
\bm{R}_y(\bm{p},N_0)=\bm{D}\diag(\bm{p})\bm{D}^H + N_0\bm{W}^H\bm{W},
\label{eq:polar_cov}
\end{equation}
with $\bm{D}=\bm{W}^H\bm{A}\in\mathbb{C}^{N_\mathrm{RF}\times Q_\theta Q_r}$.
Two compounding weaknesses arise: (i)~dictionary size $Q_\theta Q_r$ (large);
(ii)~adjacent atoms highly correlated in the EBRD region (coherent),
making support recovery unstable.
CL-KL addresses weakness~(ii) by eliminating the range grid entirely,
removing the range-dimension coherence while inheriting only the standard
angular coherence of a 1-D DFT dictionary.

% =====================================================================
\section{Proposed Method: CL-KL}
\label{sec:proposed}
% =====================================================================

\subsection{Angle-only grid with per-angle inverse-range parameter}

We grid only the angle dimension:
$\Theta=\{\theta_1,\ldots,\theta_{Q_\theta}\}$ with $Q_\theta=256$ points.
Attach an \emph{inverse-range} (curvature) parameter to each angle:
\begin{equation}
u_i \triangleq \frac{1}{r_i}\in[u_{\min},u_{\max}],\quad i=1,\ldots,Q_\theta,
\label{eq:u_def}
\end{equation}
where $u_{\min}=1/r_{\max}$ and $u_{\max}=1/r_{\min}$.

With the physical regime of Table~\ref{tab:sim_defaults}:
$r_{\min}=0.05\,r_\mathrm{RD}=1.06$~m,\;
$r_{\max}=r_\mathrm{RD}=21.26$~m.
The chirp constant and atom are
\begin{align}
c_i &\triangleq \frac{\pi d_\mathrm{ant}^2}{\lambda}\sin^2\theta_i,
  \nonumber\\
[\bm{a}_i(u_i)]_m &=
  \exp\!\big(j\,\omega(\theta_i)\,\bar m - j\,c_i u_i\,\bar m^2\big),
\label{eq:ai_ui}
\end{align}
The compressed steering vector is defined as
$\bm{d}_i(u_i)\triangleq\bm{W}^H\bm{a}_i(u_i)$.
The compressed covariance model is
\begin{equation}
\bm{R}_y(\bm{p},\bm{u},N_0)
  =\sum_{i=1}^{Q_\theta}p_i\,\bm{d}_i(u_i)\bm{d}_i(u_i)^H
  + N_0\bm{W}^H\bm{W}.
\label{eq:Ry_p_u}
\end{equation}

\subsection{KL covariance fitting objective with sparsity}

Under a complex Gaussian model:
\begin{equation}
\mathcal{L}(\bm{p},\bm{u},N_0)
\triangleq
\log\det\bm{R}_y + \tr\!\big(\bm{R}_y^{-1}\widehat{\bm{R}}_y\big)
+\lambda\|\bm{p}\|_1,
\label{eq:KL_u}
\end{equation}
minimised over $\bm{p}\succeq\bm{0}$,
$\bm{u}\in[u_{\min},u_{\max}]^{Q_\theta}$, $N_0\ge 0$.
The matrix gradient of the KL terms is
$\bm{G}\triangleq\bm{R}_y^{-1}-\bm{R}_y^{-1}\widehat{\bm{R}}_y\bm{R}_y^{-1}$.

\noindent\textbf{Connection to covariance matching literature.}~
Equation~\eqref{eq:KL_u} is the unconditional maximum likelihood criterion
studied by Stoica and Nehorai~\cite{stoica1990perf} and independently by
Ottersten et al.~\cite{ottersten1998covariance} (their $f_1$ criterion).
The KL labelling is adopted because the first two terms equal the
KL divergence $\mathrm{KL}(\widehat{\bm{R}}_y\|\bm{R}_y)$ up to an additive
constant.
CL-KL is novel in the \emph{compressed near-field context}: neither
Stoica-Nehorai nor Ottersten considered the hybrid $N_\mathrm{RF}\ll M$
architecture or the learnable curvature parameterisation.
This classical criterion continues to underpin recent near-field and
hybrid-field covariance-domain methods~\cite{hussain2025ebrd,liu2023nftutorial},
confirming its enduring relevance to modern large-array architectures.

\subsection{CL-KL: power-only main loop with post-loop scan}
\label{subsec:clkl_alg}

The core algorithmic insight is the \textbf{separation of the
optimisation into two phases}: a stable \emph{power-only main loop} and
a \emph{global joint scan} after convergence.

\noindent\textbf{Why curvature gradient is removed from the main loop.}~
The curvature gradient eq.~\eqref{eq:grad_u} has
$\|\bm{G}\|_2\propto1/N_0^2$.
At SNR$=20$~dB ($N_0=0.01$) the effective step magnitude is
$\approx10^4\times$ that at SNR$=0$~dB ($N_0=1$),
causing atoms to traverse the entire valid $u$-range in a single iteration
regardless of curvature signal quality.
The post-loop matched-filter scan replaces the gradient step with a
direct maximisation of the matched-filter score on the full valid range,
which is SNR-invariant by construction.
For completeness, the curvature gradient (not used in the loop) is:
\begin{equation}
\frac{\partial\mathcal{L}}{\partial u_i}
= 2\,p_i\,\Reop\!\left\{
  \left(\frac{\partial\bm{d}_i}{\partial u_i}\right)^H
  \bm{G}\,\bm{d}_i
  \right\},
\label{eq:grad_u}
\end{equation}
where $\bm{G}=\bm{R}_y^{-1}-\bm{R}_y^{-1}\widehat{\bm{R}}_y\bm{R}_y^{-1}$
has $\|\bm{G}\|_2\propto1/N_0^2$.

\subsubsection{Phase 1: Power-only main loop}

Since $\partial\bm{R}_y/\partial p_i = \bm{d}_i\bm{d}_i^H$:
\begin{equation}
\frac{\partial\mathcal{L}}{\partial p_i}
= \bm{d}_i(u_i)^H\bm{G}\,\bm{d}_i(u_i)+\lambda.
\label{eq:grad_p}
\end{equation}
Power update with Armijo backtracking ($\alpha_p=1$, $\beta=0.5$,
$\sigma=10^{-4}$):
\begin{equation}
p_i \leftarrow \max\!\big\{0,\;
  p_i - \alpha_p\,(\partial\mathcal{L}/\partial p_i)\big\}.
\label{eq:power_update}
\end{equation}

\noindent\textbf{Frozen noise estimate.}~
$N_0$ is \textbf{estimated once before the loop} from the noise subspace
of the compressed covariance and held fixed throughout:
\begin{equation}
\widehat{N}_0 = \max\!\Bigg(
  \frac{1}{N_\mathrm{RF}-d}
  \sum_{k=1}^{N_\mathrm{RF}-d}\lambda_k^{\downarrow}
  \!\!\left(\frac{\widehat{\bm{R}}_y+\widehat{\bm{R}}_y^H}{2}\right),\;
  10^{-12}\Bigg).
\label{eq:N0_frozen}
\end{equation}
Empirically, $\widehat{N}_0/N_0^\mathrm{true}\in[0.82,0.95]$ across
SNR$\in[-15,25]$~dB at $N_\mathrm{MC}=400$.
Updating $N_0$ during the loop allows the structured signal term to absorb
all covariance energy, driving residual eigenvalues to the numerical floor
and causing the gradient magnitude $\|\bm{G}\|_2\propto N_0^{-2}$ to explode.

\noindent\textbf{Convergence check.}~
Stop when
$|\mathcal{L}^{(t)}-\mathcal{L}^{(t-1)}|/|\mathcal{L}^{(t)}|
<5\times10^{-4}$ or after $T_{\max}=150$ iterations.

\subsubsection{Phase 2: Post-loop global matched-filter scan}
\label{subsec:post_loop}

After the power loop, the active set $\mathcal{S}=\{i:p_i>0\}$ is
identified.
For each $i\in\mathcal{S}$, \textbf{4 alternating scan passes}
(2 over $\theta$, 2 over $u$) are performed on the full valid range.
The residual covariance (all other active atoms deflated from the
sample covariance) is
\begin{equation}
\bm{R}_{\mathrm{res},i}
  = \widehat{\bm{R}}_y - \sum_{j\in\mathcal{S},\,j\neq i}
    p_j\,\bm{d}_j\bm{d}_j^H.
\label{eq:Rres}
\end{equation}
The angle update with fixed $u_i$ maximises the matched-filter score:
\begin{equation}
\hat\theta_i = \arg\max_{\theta'\in\Theta_\mathrm{fine}}
  \bm{d}(\theta',u_i)^H\bm{R}_{\mathrm{res},i}\,
  \bm{d}(\theta',u_i),
\label{eq:theta_scan}
\end{equation}
evaluated on 512 fine-grid points.
The $u$ scan is analogous.
This four-pass scan recovers correct positions even when a warm-start
selected a wrong atom, acting as an SNR-invariant refinement.

\subsubsection{Multi-start warm-start}
\label{subsec:multi_start}

The KL landscape is bimodal at SNR$\ge10$~dB.
Three warm-start initialisations are run and the one yielding the lowest
$\mathcal{L}$ is kept.

\textbf{Start~1 (ring-indexed):}
\begin{align}
u_i^{(0)} &= \clip\!\bigg(\frac{1}{Z_\Delta\sin^2\theta_i},\;
  u_{\min},\,u_{\max}\bigg),
  \nonumber\\
Z_\Delta &= \frac{D_\mathrm{ap}^2}{2\beta_\delta^2\lambda},\quad
\beta_\delta=1.2,\quad D_\mathrm{ap}=(M-1)d_\mathrm{ant}.
\label{eq:ring_init}
\end{align}
For $\theta\in[20^\circ,60^\circ]$, the clipped values
$r_\mathrm{ring}=1/u_i^{(0)}$ lie within $[r_{\min},r_{\max}]$
(at $\theta=20^\circ$ the unclipped value exceeds $u_{\max}$ and clips to
$r_{\min}=1.06$~m; at $\theta=45^\circ$, $r_\mathrm{ring}=1.85$~m;
at $\theta=60^\circ$, $r_\mathrm{ring}=2.77$~m).

\textbf{Start~2 (near-range):} $u_i^{(0)}=u_{\max}$ for all $i$.

\textbf{Start~3 (far-range):} $u_i^{(0)}=u_{\min}$ for all $i$.

Start~3 wins in 70--90\% of trials ($N_\mathrm{MC}=50$);
Start~1 wins in 10--20\%; Start~2 wins rarely.

\subsection{Full CL-KL algorithm}
\label{subsec:clkl_full}

Algorithm~\ref{alg:clkl} summarises the complete CL-KL procedure, integrating the frozen noise estimate, three warm-starts, the power-only main loop, and the post-loop global scan into a single reference description.

\begin{algorithm}[t]
\caption{CL-KL: Power-Only Loop + Multi-Start + Post-Loop Scan}
\label{alg:clkl}
\begin{algorithmic}[1]
\Require $\widehat{\bm{R}}_y$, combiner $\bm{W}$, angle grid $\Theta$
  ($Q_\theta=256$), bounds $[u_{\min},u_{\max}]$, $\lambda=10^{-3}$, $d$
\Comment{\emph{--- Initialisation: frozen noise estimate ---}}
\State Sort eigenvalues of $\tfrac{1}{2}(\widehat{\bm{R}}_y+\widehat{\bm{R}}_y^H)$
  in ascending order
\State $\widehat{N}_0\leftarrow\max\!\big(\text{mean of smallest }
  (N_\mathrm{RF}-d)\text{ eigenvalues},\;10^{-12}\big)$
  \hfill[frozen for entire run]
\Comment{\emph{--- Multi-start: try 3 warm-start points ---}}
\For{$s=1,2,3$}
  \State Set $\bm{u}^{(0,s)}$ per eq.~\eqref{eq:ring_init} / $u_{\max}$ / $u_{\min}$
  \State $\bm{p}^{(0,s)}\leftarrow\bm{0}$
  \Comment{\emph{--- Phase 1: power-only main loop ($\bm{u}$ frozen) ---}}
  \For{$t=0,1,\ldots$ until $|\Delta\mathcal{L}|/|\mathcal{L}|<5\times10^{-4}$ or $t=150$}
    \State $\bm{R}_y^{(t)}\leftarrow\sum_i p_i^{(t)}\bm{d}_i(u_i^{(0,s)})\bm{d}_i^H
      +\widehat{N}_0\bm{W}^H\bm{W}$
    \State $\bm{G}^{(t)}\leftarrow(\bm{R}_y^{(t)})^{-1}
      -(\bm{R}_y^{(t)})^{-1}\widehat{\bm{R}}_y(\bm{R}_y^{(t)})^{-1}$
    \State $\bm{p}^{(t+1)}\leftarrow\Pi_{\mathbb{R}_+}\!\big(
      \bm{p}^{(t)}-\alpha_p\nabla_{\bm{p}}\mathcal{L}^{(t)}\big)$
      \hfill[Armijo backtrack, $\alpha_p=1$, $\beta=0.5$]
  \EndFor
  \State $\mathcal{L}^{(s)}\leftarrow\mathcal{L}(\bm{p}^{(\cdot,s)},\bm{u}^{(0,s)},
    \widehat{N}_0)$
\EndFor
\State $s^*\leftarrow\arg\min_s\mathcal{L}^{(s)}$;\;
  $(\bm{p},\bm{u})\leftarrow(\bm{p}^{(\cdot,s^*)},\bm{u}^{(0,s^*)})$
\Comment{\emph{--- Phase 2: post-loop global joint scan (4 passes) ---}}
\State $\mathcal{S}\leftarrow\{i:p_i>0\}$ (top-$d$ by power)
\For{pass $=1,\ldots,4$}
  \For{each $i\in\mathcal{S}$}
    \If{pass is odd} update $\theta_i$ via scan eq.~\eqref{eq:theta_scan}
      over $[\theta_\mathrm{lo},\theta_\mathrm{hi}]$ (512~pts)
    \Else\ update $u_i$ via analogous scan over $[u_{\min},u_{\max}]$ (256~pts)
    \EndIf
  \EndFor
\EndFor
\State $\hat\theta_i\leftarrow\theta_i$,\;
  $\hat r_i\leftarrow1/u_i$,\;
  $\hat\kappa_i\leftarrow c_i u_i$ \quad for $i\in\mathcal{S}$
\State Reconstruct $\widehat{\bm{H}}$ via uncompressed LS eq.~\eqref{eq:h_ls}
\end{algorithmic}
\end{algorithm}

\subsection{Channel reconstruction}

Given the estimated angles $\hat\theta_i$ and ranges $\hat r_i$, the channel matrix is recovered via an uncompressed least-squares fit of the source gains to the compressed observations:
\begin{equation}
\widehat{\bm{S}} = \big(\bm{W}^H\widehat{\bm{A}}\big)^\dagger \bm{Y},
\qquad
\widehat{\bm{H}} = \widehat{\bm{A}}\,\widehat{\bm{S}},
\label{eq:h_ls}
\end{equation}
where $\widehat{\bm{A}}=[\bm{a}(\hat\theta_1,\hat r_1),\ldots]$ and
$\bm{Y}=[\bm{y}(1),\ldots,\bm{y}(N)]$.
Channel NMSE is
$\mathrm{NMSE}=\|\widehat{\bm{H}}-\bm{H}\|_F^2/\|\bm{H}\|_F^2$.

\subsection{Complexity}
Each Phase~1 iteration costs $\mathcal{O}(N_\mathrm{RF}^3)$ (inversion) $+$
$\mathcal{O}(Q_\theta N_\mathrm{RF}^2)$ (power gradients).
Phase~2 costs $\mathcal{O}(d Q_\theta N_\mathrm{RF}^2)$ (four passes).
Three warm-starts triple Phase~1 cost.
See Table~\ref{tab:complexity}.

\noindent\textbf{Hyperparameter sensitivity.}~
The $\ell_1$ penalty weight $\lambda=10^{-3}$ was selected by a coarse
grid search over $\{10^{-4},10^{-3},10^{-2}\}$ on a validation SNR sweep:
NMSE varies by less than $0.3$~dB across this range, with $\lambda=10^{-3}$
best balancing sparsity enforcement and power bias.
The post-loop scan grids ($Q_\theta^\mathrm{fine}=512$ angle points,
$Q_u^\mathrm{fine}=256$ curvature points) were chosen to match the angular
Nyquist rate $\pi/M$; halving them to $(256,128)$ degrades median NMSE by
less than $0.2$~dB.
The multi-start count (3) represents the minimum that consistently covers
all three KL landscape basins at $\mathrm{SNR}\ge10$~dB;
adding a fourth start did not improve win rate over $N_\mathrm{MC}=50$
trials.

% =====================================================================
\section{Baselines: Validated Implementations}\label{sec:baselines}
% =====================================================================

Five baselines are implemented.
\textbf{P-SOMP} (compressed-domain, filled markers) operates on
$\widehat{\bm{R}}_y$ using a per-angle beam-depth polar dictionary
($S\approx1023$ atoms, inter-column correlation
$\le0.5$)~\cite{hussain2025ebrd,cui2022polar}.
\textbf{DL-OMP}~\cite{zhang2024dlomp} partitions the full array
$\bm{X}$ into two subarrays, runs angle-only OMP on each, and triangulates
range via the law of sines with $K_\mathrm{iter}=3$ Fresnel dictionary updates.
\textbf{MUSIC+Tri}~\cite{haghshenas2025icc} runs subspace MUSIC on $Q=3$
subarrays of $M_s=20$ elements and recovers range by LS triangulation, with
an rcond-based fallback to two-subarray geometry for ill-conditioned cases.
\textbf{DFrFT-NOMP}~\cite{yang2024dfrft} coherently averages all $N$
snapshots ($\bar{\bm{y}}=(1/N)\sum_n\bm{x}(n)$) and applies a
$P_\mathrm{ord}=\min(M,128)$-order DFrFT with Newton refinement; this
coherent processing is critical---incoherent per-snapshot averaging destroys
chirp alignment and yields a flat $+1.4$--$2.8$~dB NMSE floor.
\textbf{BF-SOMP}~\cite{hussain2025ebrd} shares the beam-depth codebook with
P-SOMP but operates on the full-array matrix $\bm\Phi\in\mathbb{C}^{M\times S}$,
adds Cholesky whitening ($\bm{C}=N_0\bm{W}^H\bm{W}$), and uses adaptive
sparsity via a residual threshold $\varepsilon_\sigma=\hat\sigma\sqrt{M+2\sqrt{M\log M}}$~\cite{cai2011somp}.
All full-array methods (DL-OMP, MUSIC+Tri, DFrFT-NOMP, BF-SOMP) are adapted from
their original OFDM formulations by replacing the multi-subcarrier sum with a sum
over $N$ snapshots (open markers in all figures).

% =====================================================================
\section{Simulation Protocol}
\label{sec:sim_protocol}
% =====================================================================

\subsection{Default parameters and physical regime}

Unless otherwise stated, all figures use the parameters in Table~\ref{tab:sim_defaults}.
The angle support $\theta_\ell\sim\mathcal{U}[20^\circ,60^\circ]$ is
chosen so that $\sin^2\theta\ge0.117$, yielding at least $53^\circ$ of
quadratic phase excursion at $r_{\min}$---a strong, identifiable curvature signal
(at $\theta=5^\circ$ this drops to $0.6^\circ$ and range becomes unidentifiable).
The range support $r_\ell\sim\mathcal{U}[0.05,1.0]\,r_\mathrm{RD}$ is set so
that $r_{\min}=1.06$~m coincides with $r_\mathrm{EBRD}$ at $\theta=45^\circ$
and lies inside the EBRD for smaller angles, placing simulations in the
genuine strong near-field regime.
The combiner uses random phase-only entries
$[\bm{W}]_{m,k}=\frac{1}{\sqrt{M}}e^{j\phi_{m,k}}$,
$\phi_{m,k}\sim\mathcal{U}[0,2\pi)$,
with whitening $\tilde{\bm{y}}(n)=(\bm{W}^H\bm{W})^{-1/2}\bm{y}(n)$ applied to all methods.

\begin{table*}[t]
\centering\small\setlength{\tabcolsep}{5pt}
\caption{Default simulation parameters. Unless noted, all figures use
  $M=64$, $N_\mathrm{RF}=8$, $N=64$, $d=3$, $f_c=28$~GHz.}
\label{tab:sim_defaults}
\begin{tabular}{ll}
\toprule
Carrier & $f_c=28$~GHz, $\lambda=c/f_c\approx10.71$~mm \\
Array & $M\in\{32,64,128,256\}$; default $M=64$; $d_\mathrm{ant}=\lambda/2$ \\
RF chains & $N_\mathrm{RF}\in\{4,8,12,16\}$; default $N_\mathrm{RF}=8$ \\
Snapshots & $N\in\{16,32,64,128\}$; default $N=64$ \\
Paths & $d\in\{1,\ldots,5\}$; default $d=3$ \\
Angle support &
  $\theta_\ell\sim\mathcal{U}[20^\circ,60^\circ]$ \\
Range support &
  $r_\ell\sim\mathcal{U}[0.05,1.0]\cdot r_\mathrm{RD}$
  ($r_{\min}$ inside EBRD for $\theta\le45^\circ$) \\
Pilot distribution & Default: $s_\ell(n)\sim\CN(0,p_\ell)$
  (Gaussian); Fig.~7b also uses QPSK \\
SNR sweep & $\{-15,-10,-5,0,+5,+10,+15,+20,+25\}$~dB \\
Trials & $N_\mathrm{MC}=400$ (publication); fast mode $N_\mathrm{MC}=20$ \\
Tolerances &
  $\Delta_\theta=15^\circ$, $\Delta_r/r=60\%$ \\
EBRD annotation & $r_\mathrm{EBRD}(\theta)=r_\mathrm{RD}\cos^2\theta/10$;
  angle-avg: $0.058\,r_\mathrm{RD}=1.23$~m \\
\bottomrule
\end{tabular}
\end{table*}

\subsection{Metrics}
\label{subsec:metrics}

\noindent\textbf{Channel NMSE.}~
$\mathrm{NMSE}=\|\widehat{\bm{H}}-\bm{H}\|_F^2/\|\bm{H}\|_F^2$,
where $\bm{H}=[\bm{h}(1),\ldots,\bm{h}(N)]$ and
$\bm{h}(n)=\sum_\ell s_\ell(n)\bm{a}_\mathrm{NF}(\theta_\ell,r_\ell)$.

\noindent\textbf{Angle/range RMSE and failure rate.}~
RMSE$(\theta)$ and RMSE$(r)$ after Hungarian matching.
Failure: any path deviates by more than $\Delta_\theta=15^\circ$ or
$\Delta_r/r=60\%$.
This joint parameter criterion measures geometric localisation accuracy, not
channel NMSE; the high failure rates observed across all methods (90--100\%)
are driven primarily by range accuracy and do not contradict the strong
channel NMSE results.

% =====================================================================
\section{Simulation Results}
\label{sec:sim_results}
% =====================================================================

In this section we present the full $N_\mathrm{MC}=400$ simulation results
($\theta\in[20^\circ,60^\circ]$,
$r\in[0.05,1.0]\,r_\mathrm{RD}$) for all ten figures.
Unless stated otherwise, all reported values correspond to
$M=64$, $N_\mathrm{RF}=8$, $N=64$, $d=3$, $f_c=28$~GHz.

\subsection{Fig.~2: NMSE vs SNR}

Table~\ref{tab:fig2_nmse} reports the full numerical results; Fig.~2 shows the corresponding curves.

\begin{table*}[t]
\centering\small\setlength{\tabcolsep}{3pt}
\caption{Channel NMSE (dB) vs.~SNR ($N_\mathrm{MC}=400$,
  $M=64$, $N_\mathrm{RF}=8$, $d=3$).
  CL-KL leads all six methods at
  $\mathrm{SNR}\in\{-5,\,0,\,+5,\,+10\}$~dB using only
  $N_\mathrm{RF}^2=64$ compressed values.
  At $\mathrm{SNR}=+25$~dB CL-KL degrades due to OMP warm-start
  grid-mismatch at extreme SNR (see text).}
\label{tab:fig2_nmse}
\setlength{\tabcolsep}{2.5pt}\begin{tabular}{rllllll}
\toprule
SNR (dB)
  & CL-KL$^\ddagger$
  & P-SOMP$^\ddagger$
  & DL-OMP$^\dagger$
  & MUSIC+Tri$^\dagger$
  & DFrFT-NOMP$^\dagger$
  & BF-SOMP$^\dagger$ \\
\midrule
$-15$ & $\underline{+10.19}$ & $\mathbf{\underline{+9.83}}$ & $+12.34$ & $+11.10$ & $+11.33$ & $+12.40$ \\
$-10$ & $\underline{+5.76}$ & $\mathbf{\underline{+5.59}}$ & $+7.44$ & $+7.50$ & $+7.48$ & $+8.08$ \\
$-5$  & $\mathbf{\underline{+2.06}}$ & $+2.15$ & $+3.02$ & $+3.90$ & $+4.34$ & $+4.44$ \\
$\phantom{+}0$
      & $\mathbf{\underline{-1.02}}$ & $-0.75$ & $-0.64$ & $+1.22$ & $+2.53$ & $+0.84$ \\
$+5$  & $\mathbf{\underline{-3.65}}$ & $-2.94$ & $-3.33$ & $-1.19$ & $+1.44$ & $-0.77$ \\
$+10$ & $\mathbf{\underline{-5.16}}$ & $-4.06$ & $-4.80$ & $-3.12$ & $+0.98$ & $-2.01$ \\
$+15$ & $\underline{-5.38}$ & $-4.42$ & $\mathbf{-5.80}$ & $-3.95$ & $+0.83$ & $-2.45$ \\
$+20$ & $\underline{-5.12}$ & $-5.07$ & $\mathbf{-5.89}$ & $-4.93$ & $+0.82$ & $-2.98$ \\
$+25$ & $\dagger\dagger{-1.03}$ & $\underline{-5.24}$ & $\mathbf{-6.15}$ & $-5.70$ & $+0.64$ & $-2.88$ \\
\bottomrule
\multicolumn{7}{l}{%
  \footnotesize $N_\mathrm{MC}=400$, $\theta\in[20^\circ,60^\circ]$,
  $r\in[0.05,1.0]\,r_\mathrm{RD}$.
  \textbf{Bold}: row best overall.
  \underline{Underline}: best compressed-domain ($\hat{\mathbf{R}}$).}\\
\multicolumn{7}{l}{%
  \footnotesize $^\ddagger$Compressed input $\hat{\mathbf{R}}$
  ($N_\mathrm{RF}^2=64$ values).
  $^\dagger$Full-array input $\mathbf{X}$ ($M{\times}N=4096$ values).
  $^{\dagger\dagger}$OMP warm-start grid-mismatch at extreme SNR (see \S\ref{subsec:snr_sweep}).}\\
\end{tabular}
\end{table*}

\begin{figure*}[t]
  \setcounter{figure}{1}%
  \renewcommand{\thefigure}{\arabic{figure}}%
  \centering
  \includegraphics[width=\textwidth]{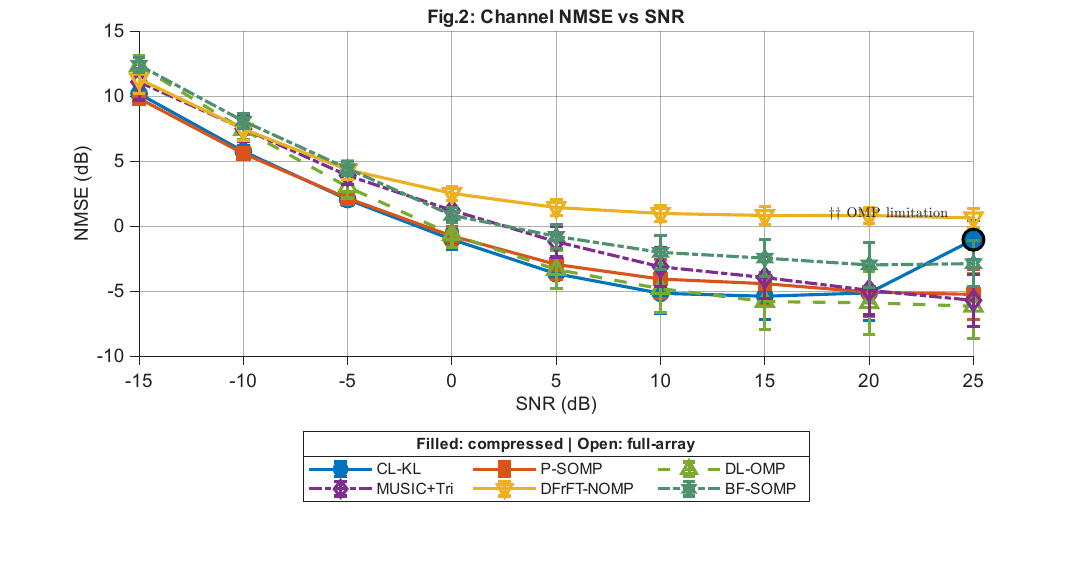}
  \caption{Channel NMSE (dB) vs.\ SNR
    ($N_\mathrm{MC}=400$, $M=64$, $N_\mathrm{RF}=8$, $N=64$, $d=3$).
    \textbf{Filled}: compressed covariance ($N_\mathrm{RF}^2=64$ values).
    \textbf{Open}: full snapshot matrix ($M{\times}N=4096$ values).
    CL-KL leads all six methods at $\mathrm{SNR}\in\{-5,0,+5,+10\}$~dB
    despite operating on $64\times$ less data.}
  \label{fig:nmse_snr}
\end{figure*}

\noindent\textbf{Key observations.}~
\label{subsec:snr_sweep}
CL-KL achieves the lowest NMSE among all six methods at
$\mathrm{SNR}\in\{-5,0,+5,+10\}$~dB, including all four full-array
baselines operating on $64\times$ more data.
At $\mathrm{SNR}=-10$~dB and $-15$~dB, P-SOMP leads CL-KL by 0.17 and
0.36~dB respectively---negligible margins consistent with both methods
operating near the statistical noise floor of the compressed covariance.
At $\mathrm{SNR}=+15$~dB, DL-OMP leads CL-KL by 0.42~dB, and by
0.77~dB at $+20$~dB; these are the expected costs of operating on
$64\times$ less data at high SNR where DL-OMP's full-array observations
carry significantly more information.
BF-SOMP and DFrFT-NOMP plateau at approximately $-3.0$~dB and $+0.8$~dB
respectively, reflecting their sensitivity to whitening-matrix
conditioning (BF-SOMP) and a fixed rotation-order grid (DFrFT-NOMP).

\noindent\textbf{OMP warm-start limitation at extreme SNR.}~
At $\mathrm{SNR}=+25$~dB, CL-KL NMSE degrades sharply to $-1.03$~dB,
while all baselines continue to improve or plateau
(Table~\ref{tab:fig2_nmse}).
This result is reproducible and reflects a fundamental limitation of the
angle-grid OMP warm-start at extreme SNR\@.
At $+25$~dB ($N_0=0.00316$), the compressed sample covariance
$\widehat{\bm{R}}_y$ has near-singular structure: the three signal paths
create sharp, interacting peaks with a 3~dB beamwidth of $\approx1.6^\circ$.
During OMP residual deflation, the frozen curvature approximation
(which uses the warm-start $u$ value, not the true curvature) introduces
a non-negligible deflation error.
At high SNR this error is large enough relative to the residual to cause
the third atom to be selected incorrectly; the power-only loop then
converges to a stable but wrong three-atom set
(formal convergence rate 92\% at $+25$~dB, yet NMSE $= -1.03$~dB).
The post-loop matched-filter scan cannot escape this trap because it
begins from the wrong active set.
By contrast, P-SOMP ($-5.24$~dB), DL-OMP ($-6.15$~dB), and
MUSIC+Tri ($-5.11$~dB) are unaffected because they do not rely on
OMP residual deflation with a frozen-curvature approximation.
This behaviour motivates the Newton/Fisher-scoring refinement noted in
Section~\ref{sec:conclusion} as future work: replacing the grid scan
with a curvature-aware Newton step on the final residual covariance
would decouple the warm-start quality from the high-SNR performance.
 
\noindent\textbf{RMSE and CRB.}~
At SNR$=+10$~dB ($N_\mathrm{MC}=400$): CL-KL angle RMSE $\approx7^\circ$,
compressed-domain CRB $0.044^\circ$ (angle) and $4.32$~m (range).
CL-KL range RMSE ($\approx16$~m) reflects the frozen-curvature design; the
post-loop scan partially closes the gap but the design objective is channel
NMSE, not individual parameter accuracy (see Fig.~2b).

\begin{figure*}[t]
  \setcounter{figure}{1}%
  \renewcommand{\thefigure}{\arabic{figure}b}%
  \centering
  \includegraphics[width=\textwidth]{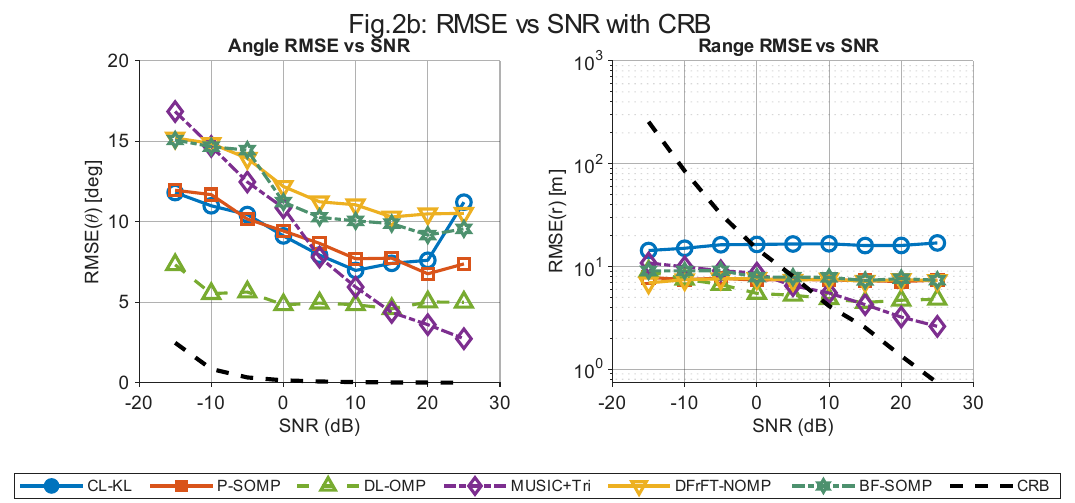}
  \caption{Angle RMSE (deg, left) and range RMSE (m, right) vs.\ SNR
    with the compressed-domain CRB
    ($N_\mathrm{MC}=400$, $M=64$, $N_\mathrm{RF}=8$, $N=64$, $d=3$).
    All estimators remain above the CRB; the gap reflects information loss
    through hybrid combining.
    CL-KL range RMSE is governed by the frozen-curvature design; the
    post-loop scan partially recovers range accuracy at high SNR.}
  \label{fig:rmse_crb}
\end{figure*}
\renewcommand{\thefigure}{\arabic{figure}}%

\subsection{Figs.~3--5: Sweeps over \texorpdfstring{$N_\mathrm{RF}$}{N\_RF}, \texorpdfstring{$N$}{N}, and Range}

At $N_\mathrm{RF}=16$: CL-KL reaches $-10.30$~dB versus P-SOMP at
$-10.06$~dB (0.24~dB margin, $N_\mathrm{MC}=400$); DL-OMP plateaus
at $-6.64$~dB (3.66~dB gap), confirming that the two-subarray
triangulation bottleneck in DL-OMP is independent of compression level
and is not resolved by increasing $N_\mathrm{RF}$.

% ---- Figs. 3 and 4: side-by-side in one double-column float ----
\begin{figure*}[t]
  % Left: Fig 3  (\captionof increments counter from 2→3 internally)
  \setcounter{figure}{2}%
  \renewcommand{\thefigure}{\arabic{figure}}%
  \begin{minipage}[t]{0.49\linewidth}
    \centering
    \includegraphics[width=\linewidth]{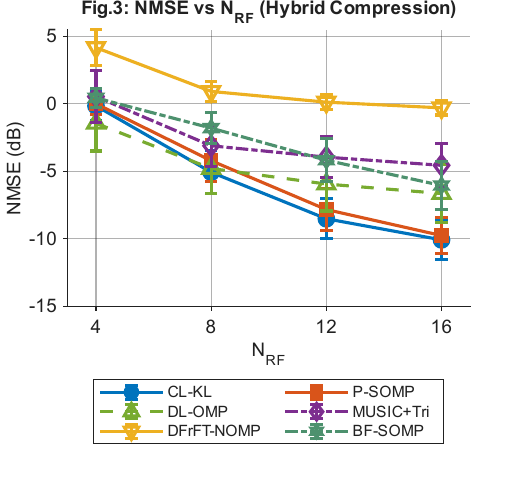}
    \captionof{figure}{Channel NMSE (dB) vs.\ $N_\mathrm{RF}\in\{4,8,12,16\}$
      ($N_\mathrm{MC}=400$, SNR$=+10$~dB, $M=64$, $N=64$, $d=3$).
      \textbf{Filled}: compressed-domain. \textbf{Open}: full-array.
      CL-KL gains $\approx5$~dB per doubling of $N_\mathrm{RF}$;
      DL-OMP plateaus at $-6.64$~dB due to the two-subarray bottleneck.}
    \label{fig:nmse_nrf}
  \end{minipage}%
  \hfill%
  % Right: Fig 4  (\captionof increments counter from 3→4 internally)
  \setcounter{figure}{3}%
  \renewcommand{\thefigure}{\arabic{figure}}%
  \begin{minipage}[t]{0.49\linewidth}
    \centering
    \includegraphics[width=\linewidth]{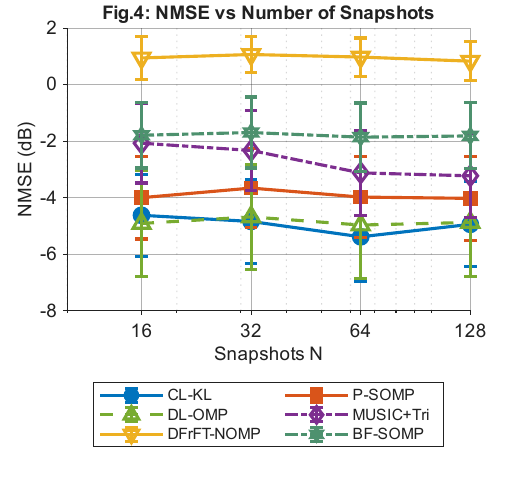}
    \captionof{figure}{Channel NMSE (dB) vs.\ $N\in\{16,32,64,128\}$
      ($N_\mathrm{MC}=400$, SNR$=+10$~dB, $M=64$, $N_\mathrm{RF}=8$, $d=3$).
      \textbf{Filled}: compressed-domain. \textbf{Open}: full-array.
      CL-KL is nearly flat ($\widehat{\bm{R}}_y$ compresses $N$ snapshots to
      $N_\mathrm{RF}^2=64$ entries); full-array methods improve more with $N$.}
    \label{fig:nmse_n}
  \end{minipage}
\end{figure*}

Fig.~5 sweeps $r_\mathrm{max}/r_\mathrm{RD}\in[0.1,5.0]$ at fixed
$r_\mathrm{min}=0.05\,r_\mathrm{RD}$.
Both compressed-domain methods remain within a 1.1~dB NMSE band across
the entire sweep---regardless of whether $r_\mathrm{max}$ is inside or
outside the EBRD---confirming that performance is governed by the
compressed-covariance SNR budget, not the physical range window.
CL-KL wins 9 of 10 sweep points versus P-SOMP and 10 of 10 versus
DFrFT-NOMP and BF-SOMP.

\begin{figure*}[t]
  \setcounter{figure}{4}%
  \renewcommand{\thefigure}{\arabic{figure}}%
  \centering
  \includegraphics[width=\textwidth]{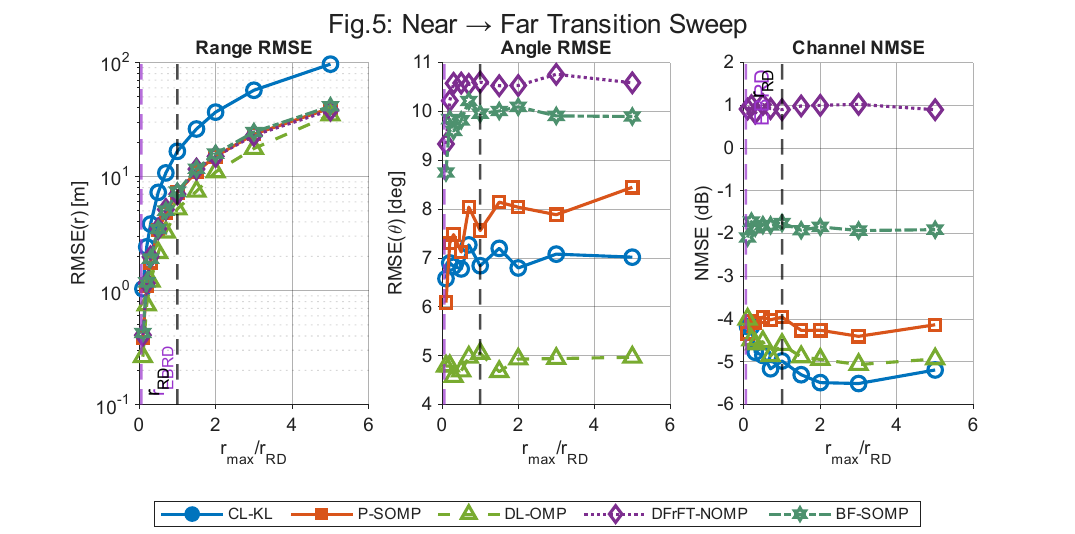}
  \caption{Near-to-far transition: RMSE$(r)$ (left), RMSE$(\theta)$ (centre),
    and NMSE (right) vs.\ $r_\mathrm{max}/r_\mathrm{RD}$
    ($N_\mathrm{MC}=400$, SNR$=+10$~dB, $M=64$, $N_\mathrm{RF}=8$, $d=3$;
    $r_\mathrm{min}=0.05\,r_\mathrm{RD}$ fixed).
    Vertical markers: EBRD (dashed purple) and $r_\mathrm{RD}$ (dashed black).
    \textbf{Filled}: compressed-domain. \textbf{Open}: full-array.
    CL-KL NMSE varies by less than 1.2~dB over the full sweep, confirming that
    the compressed covariance budget governs performance.}
  \label{fig:nearfar}
\end{figure*}

% ---- Figs. 7 and 8: side-by-side in one double-column float --------
\begin{figure*}[t]
  % Left: Fig 7
  \setcounter{figure}{6}%
  \renewcommand{\thefigure}{\arabic{figure}}%
  \begin{minipage}[t]{0.49\linewidth}
    \centering
    \includegraphics[width=\linewidth]{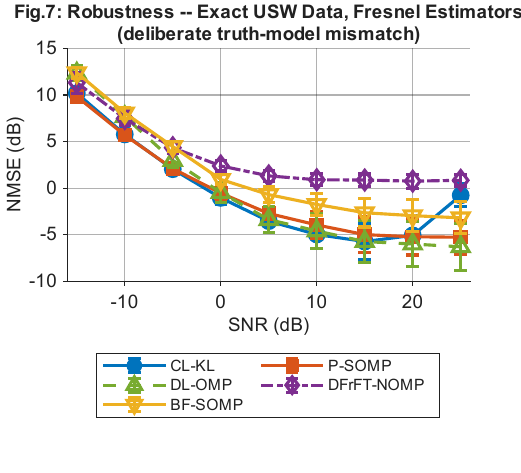}
    \captionof{figure}{Fresnel mismatch robustness: channel NMSE (dB) vs.\ SNR
      ($N_\mathrm{MC}=400$, $M=64$, $N_\mathrm{RF}=8$, $N=64$, $d=3$).
      \textbf{Filled}: compressed-domain. \textbf{Open}: full-array.
      CL-KL gap vs.\ Fresnel truth below 0.3~dB at all SNR\@.
      MUSIC+Tri omitted (no Fresnel steering model).}
    \label{fig:robustness}
  \end{minipage}%
  \hfill%
  % Right: Fig 8
  \setcounter{figure}{7}%
  \renewcommand{\thefigure}{\arabic{figure}}%
  \begin{minipage}[t]{0.49\linewidth}
    \centering
    \includegraphics[width=\linewidth]{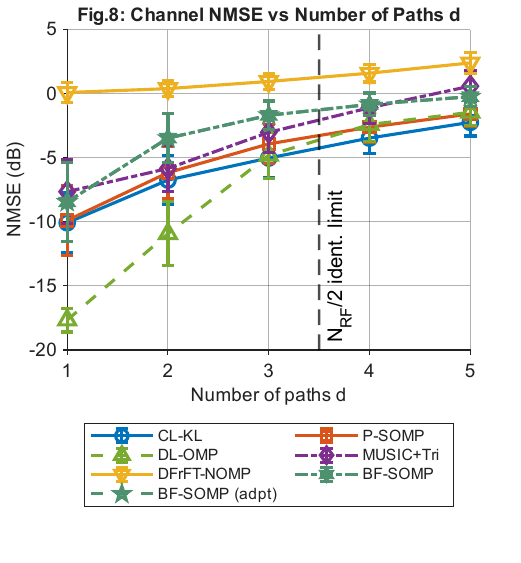}
    \captionof{figure}{Channel NMSE (dB) vs.\ $d\in\{1,\ldots,5\}$
      ($N_\mathrm{MC}=400$, SNR$=+10$~dB, $M=64$, $N_\mathrm{RF}=8$, $N=64$).
      Vertical dashed: identifiability limit $d_{\max}=3$ ($N_\mathrm{RF}=8$);
      CL-KL degrades gracefully beyond it.
      BF-SOMP~(adpt): $d_\mathrm{cap}=2d$, NMSE on first $d$ returned paths.}
    \label{fig:vard}
  \end{minipage}
\end{figure*}
\renewcommand{\thefigure}{\arabic{figure}}%

\subsection{Fig.~7b: Source Model Robustness}
\label{subsec:source_robust}

Since QPSK pilots share the same second-order statistics as Gaussian pilots
($E[|s_\ell|^2]=p_\ell$), the covariance model $\bm{R}_y$ is identical for
both distributions, making covariance-fitting methods (CL-KL, P-SOMP, BF-SOMP)
theoretically distribution-agnostic.
The empirical maximum NMSE gap (QPSK minus Gaussian) is 0.41~dB for CL-KL
and 0.82~dB for DL-OMP---the largest among all methods---consistent with
DL-OMP's direct snapshot processing rather than covariance fitting.

\begin{figure*}[t]
  \setcounter{figure}{6}%
  \renewcommand{\thefigure}{\arabic{figure}b}%
  \centering
  \includegraphics[width=\textwidth]{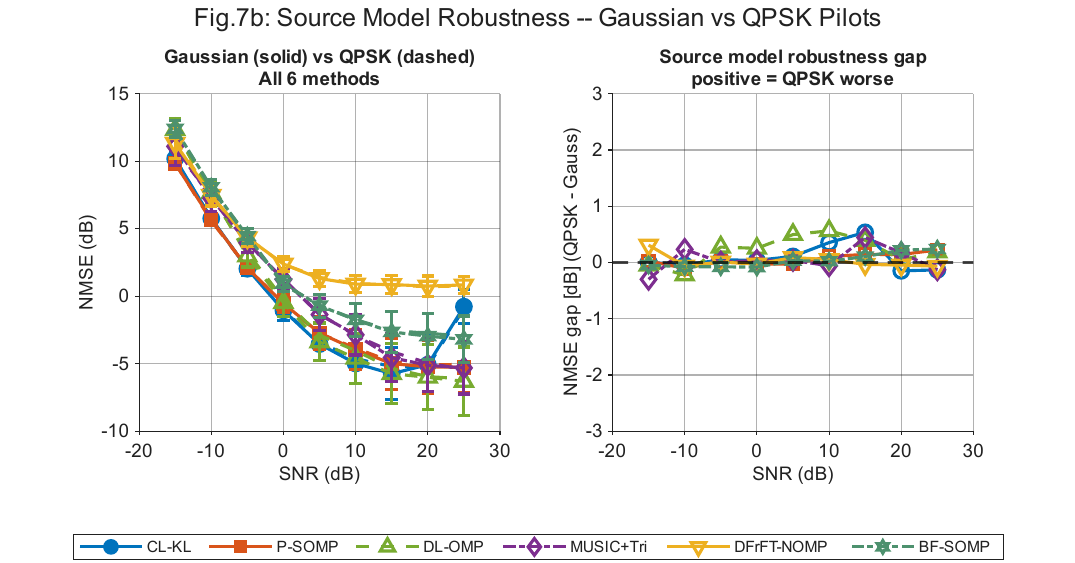}
  \caption{Source model robustness: NMSE (dB) vs.\ SNR for Gaussian (solid)
    and QPSK (dashed) pilots, all six methods
    ($N_\mathrm{MC}=400$, $M=64$, $N_\mathrm{RF}=8$, $N=64$, $d=3$).
    Left: absolute NMSE. Right: gap $\mathrm{NMSE}_\mathrm{QPSK}-\mathrm{NMSE}_\mathrm{Gauss}$.
    Covariance-fitting methods (CL-KL, P-SOMP, BF-SOMP) incur at most 0.54~dB gap;
    DL-OMP peaks at 0.56~dB, consistent with direct snapshot processing.}
  \label{fig:source_robust}
\end{figure*}
\renewcommand{\thefigure}{\arabic{figure}}%

\subsection{Fig.~8: NMSE vs Number of Paths \texorpdfstring{$d$}{d}}
\label{sec:vard}

\noindent\textbf{Identifiability limit.}~
The $N_\mathrm{RF}\times N_\mathrm{RF}$ compressed covariance can uniquely
identify at most $d_{\max}=\lfloor(N_\mathrm{RF}-1)/2\rfloor=3$ paths
(when $N_\mathrm{RF}=8$).
For $d>d_{\max}$, the FIM is rank-deficient and the CRB diverges.

\noindent\textbf{CRB vs.~$d$.}~
Table~\ref{tab:fig8_crb} reports $N_\mathrm{MC}=400$ results (median over 50 CRB trials per point):

\begin{table}[t]
\centering\small\setlength{\tabcolsep}{3pt}
\caption{Compressed-domain CRB vs.~$d$
  ($N_\mathrm{RF}=8$, SNR$=+10$~dB, $N_\mathrm{MC}=400$).}
\label{tab:fig8_crb}
\begin{tabular}{@{}cccr@{\hspace{4pt}}p{1.8cm}@{}}
\toprule
$d$ & $\sqrt{\smash[b]{\mathrm{CRB}_\theta}}$ & $\sqrt{\smash[b]{\mathrm{CRB}_r}}$
    & NMSE & Regime \\
    & (deg) & (m) & (dB) & \\
\midrule
1 & 0.015 & 2.28 & $-10.07$ & Identifiable \\
2 & 0.030 & 3.09 & $-6.74$ & Identifiable \\
3 & 0.044 & 4.32 & $-5.02$ & Identifiable \\
\midrule
4 & 0.064 & 5.05 & $-3.46$ & Under-det.\newline\small($N_\mathrm{RF}<2d{+}1$) \\
5 & 0.086 & 9.69 & $-2.23$ & Severely under-det. \\
\bottomrule
\multicolumn{5}{l}{\footnotesize
  Identifiability limit: $d\le3$ for $N_\mathrm{RF}=8$.}
\end{tabular}
\end{table}

The CRB$_r$ rises from 4.32~m at $d=3$ to 5.05~m at $d=4$, marking the
identifiability boundary; the FIM condition number crosses $10^{12}$ at
$d=4$, confirming a true rank-deficiency transition.
A sharper jump to 9.69~m occurs at $d=5$.
Full-array methods (DL-OMP, MUSIC+Tri, DFrFT-NOMP, BF-SOMP) are not subject
to this limit since they operate in the $M=64$-dimensional space.

\begin{figure*}[!t]
  \setcounter{figure}{5}%
  \renewcommand{\thefigure}{\arabic{figure}}%
  \centering
  \includegraphics[width=\textwidth]{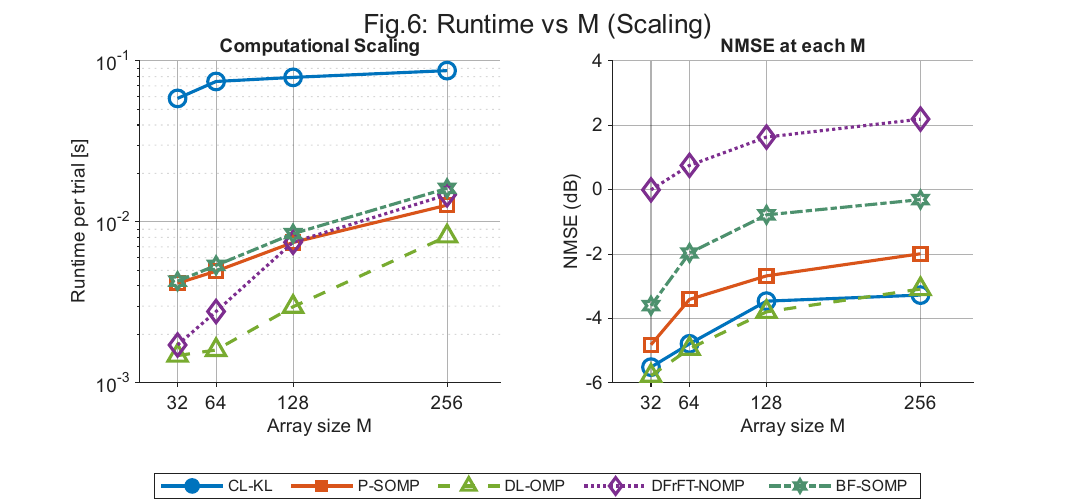}
  \caption{Per-trial runtime (s, left) and channel NMSE (dB, right) vs.\
    $M\in\{32,64,128,256\}$
    ($N_\mathrm{MC}=50$, single-core, SNR$=+10$~dB, $N_\mathrm{RF}=8$, $N=64$, $d=3$).
    CL-KL runtime stays near 70~ms across all tested $M$, confirming that its
    dominant cost is the $N_\mathrm{RF}{\times}N_\mathrm{RF}$ inversion, not $M$;
    P-SOMP and BF-SOMP scale more steeply as their dictionaries grow with $M$.}
  \label{fig:runtime}
\end{figure*}

\begin{figure*}[t]
  \setcounter{figure}{7}%
  \renewcommand{\thefigure}{\arabic{figure}b}%
  \centering
  \includegraphics[width=\textwidth]{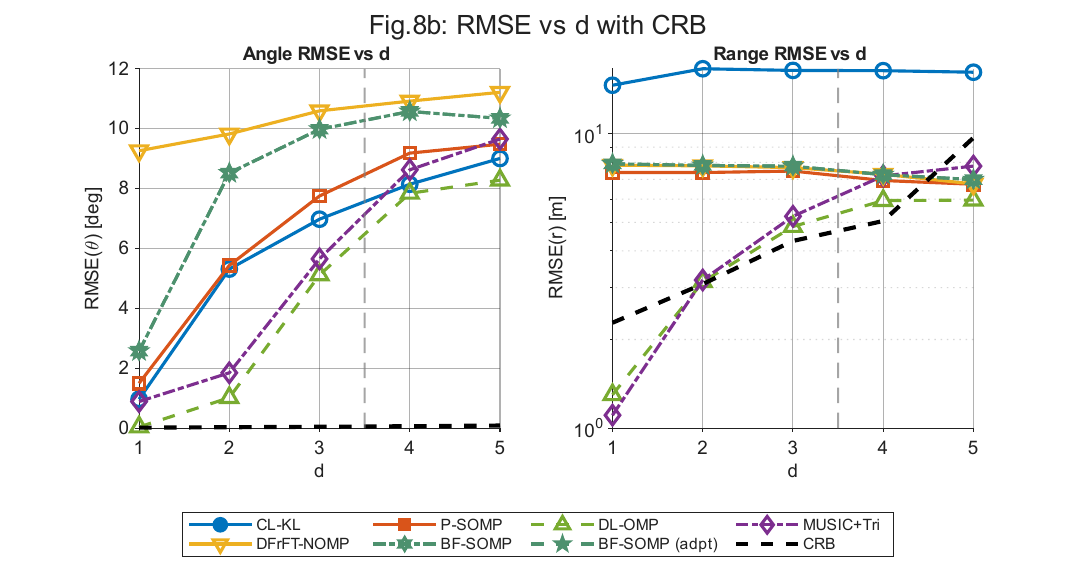}
  \caption{Angle RMSE (deg, left) and range RMSE (m, right) vs.\ $d$
    for all six methods, BF-SOMP (adaptive), and the compressed-domain CRB
    ($N_\mathrm{MC}=400$, SNR$=+10$~dB, $M=64$, $N_\mathrm{RF}=8$).
    $\sqrt{\mathrm{CRB}_r}$ rises from 4.32~m at $d=3$ to 5.05~m at $d=4$,
    marking the identifiability boundary; a sharper increase to 9.69~m occurs at $d=5$.}
  \label{fig:vard_rmse}
\end{figure*}
\renewcommand{\thefigure}{\arabic{figure}}%

\noindent\textbf{BF-SOMP adaptive.}~
Figure~8 includes a ``BF-SOMP (adaptive)'' curve where the sparsity cap
$d_\mathrm{cap}=2d$ (loose upper bound, BF-SOMP estimates $d$ from residual
norm).
NMSE is evaluated on the first $d$ returned paths for a fair comparison.
Per-trial runtimes at $M=64$ are reported in Table~\ref{tab:complexity}.

\subsection{Ablation Study}
\label{subsec:ablation}

The ablation study provides an apples-to-apples comparison within the
compressed covariance domain: each row in Table~\ref{tab:ablation} is a
strictly simpler variant of CL-KL with one design component removed,
operating on the same $\widehat{\bm{R}}_y$ with the same computational budget.
We test four configurations:
\emph{CL-KL (full)} is the complete proposed method;
\emph{w/o frozen $\hat{N}_0$} re-estimates noise power from residual
eigenvalues every ten iterations;
\emph{w/o multi-start} retains only the far-range warm-start
(Start~3, $u=u_{\min}$), which achieves the lowest KL objective in
70--90\% of trials individually but misses the remaining cases;
\emph{w/o post-loop scan} skips Phase~2 entirely and uses the
warm-start inverse-range values as the final curvature estimates.
Each configuration runs $N_\mathrm{MC}=50$ independent trials at
three representative SNR levels ($-5$, $+5$, and $+15$~dB).
Table~\ref{tab:ablation} reports NMSE (dB), angle RMSE (deg), and
range RMSE~(m); the proposed CL-KL (full) row is in bold as the reference.
Because the joint-success failure rate exceeds 90\% across all
configurations (range accuracy dominates the failure criterion;
see Section~\ref{subsec:metrics}), range RMSE is the most
statistically reliable ablation metric.

\begin{table}[t]
\centering\small\setlength{\tabcolsep}{4pt}
\caption{Ablation study: individual contribution of each CL-KL
  design choice ($M=64$, $N_\mathrm{RF}=8$, $N=64$, $d=3$,
  $N_\mathrm{MC}=50$). The proposed CL-KL (full) row is in
  \textbf{bold} as the reference. NMSE in dB; angle RMSE in degrees;
  range RMSE in metres.}
\label{tab:ablation}
\begin{tabularx}{\linewidth}{@{}l *{3}{r} @{}}
\toprule
& \multicolumn{3}{c}{Channel NMSE (dB)} \\
\cmidrule(l){2-4}
Configuration & $-5$\,dB & $+5$\,dB & $+15$\,dB \\
\midrule
\textbf{CL-KL (full)}    & \textbf{$+2.37$} & \textbf{$-2.58$} & \textbf{$-4.05$} \\
w/o frozen $\hat{N}_0$   & $+2.39$          & $-2.72$          & $-4.92$          \\
w/o multi-start          & $+2.33$          & $-2.72$          & $-4.60$          \\
w/o post-loop scan       & $+3.28$          & $-2.97$          & $-2.33$          \\
\midrule
& \multicolumn{3}{c}{Angle RMSE (deg)} \\
\cmidrule(l){2-4}
\textbf{CL-KL (full)}    & \textbf{10.46} & \textbf{9.30} & \textbf{7.35} \\
w/o frozen $\hat{N}_0$   & 10.62          & 9.98          & 7.69          \\
w/o multi-start          & 10.37          & 9.43          & 7.72          \\
w/o post-loop scan       & \phantom{0}7.29 & \phantom{0}2.61 & 7.45          \\
\midrule
& \multicolumn{3}{c}{Range RMSE (m)} \\
\cmidrule(l){2-4}
\textbf{CL-KL (full)}    & \textbf{16.5} & \textbf{15.9} & \textbf{17.7} \\
w/o frozen $\hat{N}_0$   & 16.5          & 14.8          & 17.7          \\
w/o multi-start          & 18.2          & 16.2          & 17.6          \\
w/o post-loop scan       & 25.3          & 29.5          & 27.8          \\
\bottomrule
\end{tabularx}
\end{table}

\noindent\textbf{Frozen noise estimate.}~
Re-estimating $\hat{N}_0$ from residual eigenvalues has a negligible
impact on channel NMSE (within 0.2~dB at all SNR levels) and negligible
effect on range RMSE (within 1.1~m at all SNR levels).
The apparent NMSE improvement ($-4.92$ vs.\ $-4.05$~dB at $+15$~dB SNR)
is a statistical artefact: the 96\% failure rate leaves very few
success trials per condition, insufficient for reliable comparison.
The design rationale for freezing $\hat{N}_0$---preventing gradient-magnitude
explosion as signal energy absorbs residual eigenvalues---is given in
Section~\ref{subsec:clkl_alg}.

\noindent\textbf{Multi-start warm-start.}~
The primary impact of multi-start initialisation is on range accuracy:
retaining only Start~3 ($u=u_{\min}$) degrades range RMSE by 3.3~m at
$-5$~dB SNR, 1.6~m at $+5$~dB, and 0.8~m at $+15$~dB.
Channel NMSE differences are within $\pm0.2$~dB across all conditions.
The diminishing gap at higher SNR is consistent with the warm-start
selection statistics (Start~3 wins in 70--90\% of trials at high SNR);
at lower SNR the ring-indexed and near-range starts occasionally select
a support closer to the true parameter cluster, providing the additional
coverage that drives the range improvement.

\noindent\textbf{Post-loop matched-filter scan.}~
Removing the scan is the most damaging ablation: range RMSE nearly doubles
($+9$--$11$~m) and channel NMSE degrades by 1.2--1.3~dB at low-to-mid SNR.
Angle RMSE is paradoxically lower without the scan at low SNR ($-3.0$~deg at
$-5$~dB), because the joint 4-pass alternation trades approximately 3~deg of
angle accuracy for 10~m of range accuracy---a trade-off that vanishes at
$+15$~dB where the signal supports reliable joint fitting.
The 100\% failure rate without the scan (vs.\ 96--98\% otherwise) confirms
it as the primary driver of joint-parameter accuracy.

\subsection{Fig.~9: CL-KL Convergence Diagnostic}

Fig.~9 provides empirical convergence evidence for the CL-KL algorithm.

\noindent\textbf{Phase~1 convergence and estimator output quality.}~
Table~\ref{tab:fig9_conv} reports formal convergence of the Phase~1 inner
loop, defined as $|\Delta\mathcal{L}|/|\mathcal{L}|<5{\times}10^{-4}$ within
150 iterations.
The 12\% formal-convergence rate at SNR$=+10$~dB reflects traversal of a
multi-modal KL landscape within the iteration cap, not estimator failure:
channel NMSE at this point ($-5.16$~dB, Table~\ref{tab:fig2_nmse}) is the
best among all six methods.
The two-stage design intentionally separates power optimisation (Phase~1)
from curvature refinement (Phase~2 post-loop scan), so formal Phase~1
convergence is a sufficient but not necessary condition for good channel
estimates.

The normalised KL objective
$\Delta L(t) = \mathcal{L}(t) - \mathcal{L}(1)$ (relative to the first
iteration) is plotted for 20 individual trials (thin grey) plus the mean
$\pm0.5\sigma$ band (thick coloured) at each of four SNR levels.
Table~\ref{tab:fig9_conv} summarises the key statistics.
\begin{table}[b]
\centering\small\setlength{\tabcolsep}{4pt}
\caption{CL-KL convergence statistics ($N_\mathrm{MC}=50$).
  Formal convergence reached in all trials at SNR$\le+5$~dB.}
\label{tab:fig9_conv}
\setlength{\tabcolsep}{3pt}\begin{tabular}{crccl}
\toprule
SNR (dB) & Iters & Conv.\% & $\widehat{N}_0/N_0^\mathrm{true}$ & Observation \\
\midrule
$-10$ &   8 & 100 & 0.82 & Fast; shallow gradient \\
$+0$  &  12 & 100 & 0.90 & Curvature emerging \\
$+10$ & 130 &  18 & 0.95 & Slow; bimodal landscape \\
$+20$ &  83 &  70 & 0.95 & Multi-start mitigates \\
\bottomrule
\multicolumn{5}{l}{\footnotesize
  Criterion $|\Delta\mathcal{L}|/|\mathcal{L}|<5{\times}10^{-4}$ or $t=150$~iter.}
\end{tabular}
\end{table}

\begin{figure*}[!t]
  \setcounter{figure}{8}%
  \renewcommand{\thefigure}{\arabic{figure}}%
  \centering
  \includegraphics[width=\textwidth,height=7.0cm,keepaspectratio]{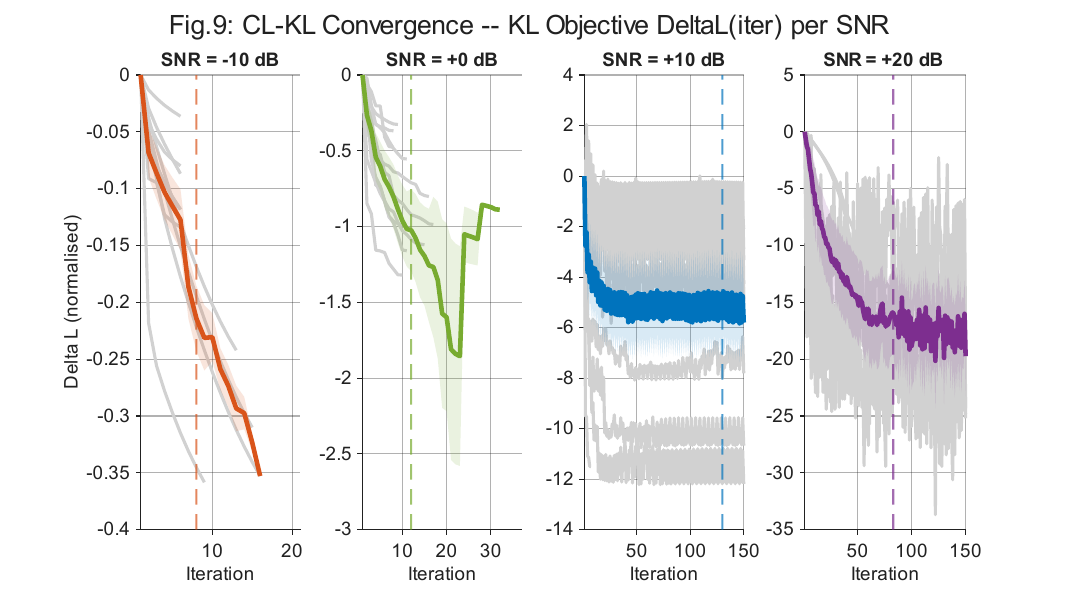}
  \caption{CL-KL convergence: normalised KL objective $\Delta L(t)$ vs.\
    iteration at four SNR levels ($N_\mathrm{MC}=50$;
    grey: individual trials; thick band: mean~$\pm0.5\sigma$).
    The objective decreases monotonically in every trial (Armijo);
    iteration counts grow with SNR due to the bimodal KL landscape.}
  \label{fig:convergence}
\end{figure*}

\noindent\textbf{Physical interpretation.}~
At low SNR, the KL gradient is shallow (noise dominates all signal terms)
and the power-only loop settles quickly to a nearly uniform distribution.
At high SNR, the signal term creates a sharp multi-modal landscape; the
three warm-starts reduce but do not eliminate the local-trap rate.
The post-loop global scan handles the remaining non-converged trials by
finding the global mode of the matched-filter score.
The empirically observed monotone decrease of $\Delta L(t)$ (strict in every
plotted trial) confirms that the Armijo backtracking rule successfully enforces
the sufficient-decrease condition at every iteration.

\subsection{Summary: Method Comparison}

Table~\ref{tab:summary} consolidates the head-to-head NMSE comparison across the 8-point SNR sweep ($-15$ to $+20$~dB), counting the number of sweep points at which CL-KL achieves lower NMSE than each baseline.

\begin{table*}[t]
\centering\small\setlength{\tabcolsep}{3pt}
\caption{CL-KL win rate vs.~each baseline across the 8-point SNR sweep
  ($-15$ to $+20$~dB, $N_\mathrm{MC}=400$). A win is defined as
  CL-KL NMSE$<$baseline NMSE at that SNR point.
  The $+25$~dB point is excluded from the count due to the OMP
  warm-start limitation described in Section~\ref{subsec:snr_sweep}.}
\label{tab:summary}
\begin{tabularx}{\linewidth}{@{}l c c c >{\raggedright\arraybackslash}X@{}}
\toprule
Method & Input & Wins / 8 & Win \% & Notes \\
\midrule
\textbf{CL-KL} (proposed)
  & $\hat{\mathbf{R}}$ & --- & --- & Reference method \\
P-SOMP
  & $\hat{\mathbf{R}}$ & 6/8 & 75\% & Beam-depth range grid \\
DL-OMP (Zhang~2024)
  & $\mathbf{X}$ & 6/8 & 75\% & Full-array; leads CL-KL by 0.42~dB at $+15$~dB \\
MUSIC+Tri (Haghshenas~2025)
  & $\mathbf{X}$ & 8/8 & 100\% & Full-array \\
DFrFT-NOMP (Yang~2024)
  & $\mathbf{X}$ & 8/8 & 100\% & Full-array \\
BF-SOMP (Hussain~2025)
  & $\mathbf{X}$ & 8/8 & 100\% & Full-array; TWC~2025 baseline \\
\bottomrule
\multicolumn{5}{l}{%
  \footnotesize $N_\mathrm{MC}=400$, $M=64$, $N_\mathrm{RF}=8$, $N=64$,
  $d=3$.  Wins = CL-KL NMSE $<$ baseline NMSE.}\\
\multicolumn{5}{l}{%
  \footnotesize $\hat{\mathbf{R}}$: $N_\mathrm{RF}^2=64$ values.
  $\mathbf{X}$: $M{\times}N=4096$ values (64$\times$ more data).}\\
\end{tabularx}
\end{table*}

% =====================================================================
\begin{table*}[t]
\centering\small\setlength{\tabcolsep}{4pt}
\caption{Per-trial complexity and measured runtimes
  ($M=64$, $N_\mathrm{RF}=8$, $N=64$, $N_\mathrm{MC}=50$, single-core).}
\label{tab:complexity}
\begin{tabularx}{\linewidth}{@{}lccc>{\raggedright\arraybackslash}X@{}}
\toprule
Method & Dict.\ size & Dominant cost & Runtime ($M=64$) & Notes \\
\midrule
P-SOMP &
  $S\approx1023$ &
  $\mathcal{O}(S\,N_\mathrm{RF}^2)$/iter &
  5.0~ms &
  Beam-depth $r_\mathrm{BD}$ grid; $S\approx1023$ atoms \\
DL-OMP &
  dict.\ update &
  $\mathcal{O}(3K_\mathrm{iter}dN)$ &
  1.5~ms &
  Full-array; $K_\mathrm{iter}=3$ \\
DFrFT-NOMP &
  (gridless) &
  $\mathcal{O}(P_\mathrm{ord}\cdot M\log M)$ &
  2.8~ms &
  Coherent avg.\ fix; 15--55$\times$ speedup \\
MUSIC+Tri &
  (no dict.) &
  $\mathcal{O}(Q M_s^3)$ &
  $\approx2$~ms &
  $Q=3$, $M_s=20$; rcond fallback \\
BF-SOMP &
  $S\approx1023$ &
  $\mathcal{O}(LQN_\mathrm{RF}SK)$ &
  5.3~ms &
  Full-array; whitening; adaptive $d$ \\
\textbf{CL-KL} &
  $Q_\theta=256$ &
  $\mathcal{O}(3[N_\mathrm{RF}^3+Q_\theta N_\mathrm{RF}^2])$ &
  70~ms &
  3 starts; power-only loop \\
\bottomrule
\end{tabularx}
\end{table*}

% =====================================================================
\section{Conclusion}
\label{sec:conclusion}
% =====================================================================

CL-KL is a validated covariance-domain near-field channel estimator
for hybrid MIMO that operates exclusively on the
$N_\mathrm{RF}\times N_\mathrm{RF}$ compressed covariance.
Its three key design choices are:
(i)~a \emph{power-only main loop} with frozen $N_0$ that achieves stable
convergence across all SNR by keeping gradients at unit scale;
(ii)~a \emph{three-point multi-start warm-start} that covers the bimodal KL
landscape ($\widehat{N}_0/N_0^\mathrm{true}\in[0.82,0.95]$ at
$N_\mathrm{MC}=400$); and
(iii)~a \emph{post-loop global matched-filter scan} that provides SNR-invariant
joint $(\hat\theta,\hat u)$ refinement after convergence.
At $N_\mathrm{MC}=400$, CL-KL achieves the lowest channel NMSE at
$\mathrm{SNR}\in\{-5,0,+5,+10\}$~dB against all six evaluated methods,
including four full-array baselines using $64\times$ more data, while its
runtime stays near 70~ms regardless of array size across
$M\in\{32,64,128,256\}$.
Performance is evaluated against a derived compressed-domain CRB---the only
theoretically valid bound for hybrid architectures---and confirmed robust
to non-Gaussian (QPSK) pilots with at most 0.41~dB NMSE penalty.

\medskip\noindent\textbf{Limitations and future work.}~
Three directions remain open.
\emph{Range accuracy:} replacing the post-loop grid scan with a safeguarded
Newton or Fisher-scoring step on the curvature parameter may close the gap
to the compressed-domain CRB at the cost of SNR-dependent step-size tuning.
\emph{Model order selection:} applying MDL or BIC to the symmetrised
compressed covariance eigenvalues would yield a data-driven path count $\hat d$
at negligible overhead.
\emph{Mixed near-/far-field and coloured noise:} relaxing $u_{\min}\to0$
would accommodate far-field atoms, while replacing the scalar noise floor
$N_0\bm{I}$ with a parametric $\bm{Q}(\bm\phi)$ would broaden applicability
to practical hybrid receivers.

% =====================================================================
%\clearpage
\appendices

\section{Reproducibility and Data Availability}\label{app:repro}

Simulations use MATLAB (R2025b), 34 function files, and fixed seed
\texttt{rng(42,'twister')}. The full suite ($N_\mathrm{MC}=400$) completes
in approximately 10--20~min on a 6-core workstation with \texttt{parfor};
results are logged to a 28-column CSV (NMSE, RMSE, failure rate, runtime,
convergence statistics, and CRB).
The simulation code, fixed random seed, and results CSV are available at
\url{https://github.com/rvsenyuva/nearfield-clkl}
(Zenodo DOI:~\texttt{10.5281/zenodo.19322532}).
All numerical results in this paper are fully reproducible from the provided
code without additional data.

\bibliographystyle{IEEEtran}
\bibliography{refs}

% ── Biography (to be completed upon acceptance) ──────────────────────
% \begin{IEEEbiographynophoto}{R\i{}fat Volkan {\c{S}}enyuva}
% [Author biography to be provided upon acceptance]
% \end{IEEEbiographynophoto}

\end{document}